\documentclass[twocolumn,prl,sort&compress,floatfix]{revtex4-1}

\usepackage{amssymb}
\usepackage{amsmath}
\usepackage{mathrsfs}
\usepackage{bm}
\usepackage[dvipsnames]{xcolor}
\usepackage[sort&compress]{natbib}

\usepackage{graphicx}

\usepackage{color}
\definecolor{red}{rgb}{1,0,0}
\definecolor{blue}{rgb}{0,0,1}

\makeatletter

\let\c@table\c@figure
\makeatother

\begin{document}

\title{Integrating Transposable Elements in the 3D Genome}

\author{Alexandros Bousios$^a$, Hans-Wilhelm Nuetzmann$^b$, Dorothy Buck$^{c}$, Davide Michieletto$^{c,d,e}$}
\affiliation{$^a$ School of Life Sciences, University of Sussex, UK}
\affiliation{$^b$ Milner Centre for Evolution, Department of Biology and Biochemistry, University of Bath, North Rd, Bath BA2 7AY, UK}
\affiliation{$^c$ Centre for Mathematical Biology and Department of Mathematical Sciences, University of Bath, North Rd, Bath BA2 7AY, UK}
\affiliation{$^d$ School of Physics and Astronomy, University of Edinburgh, Edinburgh EH9 3FD, UK}
\affiliation{$^e$ MRC Human Genetics Unit, Institute of Genetics and Molecular Medicine, University of Edinburgh, Edinburgh EH4 2XU, UK}

\begin{abstract}
	\textbf{
Chromosome organisation is increasingly recognised as an essential component of genome regulation, cell fate and cell health. Within the realm of transposable elements (TEs) however, the spatial information of how genomes are folded is still only rarely integrated in experimental studies or accounted for in modelling. Here, we propose a new predictive modelling framework for the study of the integration patterns of TEs based on extensions of widely employed polymer models for genome organisation. Whilst polymer physics is  recognised as an important tool to understand the mechanisms of genome folding, we now show that it can also offer orthogonal and generic insights into the integration and distribution profiles (or ``topography'') of TEs across organisms. Here, we present polymer physics arguments and molecular dynamics simulations on TEs inserting into heterogeneously flexible polymers and show with a simple model that polymer folding and local flexibility affects TE integration patterns. The preliminary discussion presented herein lay the foundations for a large-scale analysis of TE integration dynamics and topography as a function of the three-dimensional host genome.
} 
\end{abstract}

\maketitle
 
\section{Background}
Transposable elements (TEs) are DNA sequences that can move from one location of the genome to another. By being able to spread their own DNA across the genome independent of the cell's replication cycle~\cite{Deniz2019}, TEs represent the majority of genomic content in most eukaryotes. For example, they comprise 85 \% of the maize genome~\cite{Schnable2009} and up to 50 \% of primate genomes~\cite{Lee2016}. As such, TE activity is a major driver of phenotypic and genotypic evolution~\cite{Chuong2017,Bousios2016} and affects key biological processes from meiosis and transcription to immunological responses~\cite{Schatz2011}. At the same time, TEs have been associated with various diseases and cancer in humans~\cite{Payer2019}.

Most TEs transpose via cut-and-paste or copy-and-paste mechanisms that can both result in a net increase of the TE copy number~\cite{Wicker2007}. Amplification phases, or bursts, of TEs can occur multiple times in the evolutionary history of the host and may produce hundreds if not thousands of new copies within short time windows~\cite{Lu2017,Sanchez2017,Bousios2016}. 

Most TEs exhibit some level of integration site selection, from very specific target sites~\cite{Eickbush2014} to non-random but more dispersed genomic biases~\cite{Shinn2002,Craigie2012,Kvaratskhelia2014}. Short DNA motifs, epigenetic marks and nuclear proteins have been associated with such integration site preferences. For example, yeast Ty1 retrotransposons integrate upstream of Pol III-transcribed genes through a direct interaction between the integrase complex and the AC40 subunit of Pol III~\cite{Cheung2016}. In contrast, in plants and fungi, the integrase of certain Gypsy retrotransposons contains a chromodomain that can bind to repressive histone marks and aid insertion into heterochromatin~\cite{Gao2008}.

While the role of protein tethering and DNA motifs in TE integration is well established by now, it remains elusive how the three-dimensional (3D) structure of chromosomes and the nuclear environment is affecting TE spreading in host genomes. Chromosome folding and nuclear organisation have been shown to play key roles in all major DNA related processes~\cite{Lieberman-Aiden2009,Dekker2017,Nagano2013,Kind2013,Flyamer2017,Nozawa2017,Dong2017,Stadhouders2019}, from transcription and replication to DNA repair, and it appears only natural to expect that transposition will also be affected by the 3D organisation genome. 

A number of recent reports have highlighted that TE activity is involved in shaping 3D chromosome structure. For example, TEs of diverse families have been implicated in the establishment and maintenance of insulator boundaries between so-called ``topologically associated domains'' (TADs)~\cite{Dixon2012,Wang2015,Raviram2018,Zhang2019,Kruse2019,Sun2018}. Furthermore, TE amplification is suggested to account for a significant amount of binding motifs for the CTCF~\cite{Kunarso2010,Schmidt2012,Choudhary2018} protein, a key regulator of 3D chromosome organisation~\cite{Nora2017}. The involvement of TEs in the establishment of evolutionarily conserved long-range chromosomal interactions has been shown in different organisms~\cite{Cournac2016,Winter2018} and some of these TE-mediated interactions appear to be of functional importance in gene regulation. In the plant Arabidopsis thaliana, TEs are enriched at genomic hubs of long-range chromosomal interactions with anticipated functional roles in silencing of foreign DNA elements~\cite{Grob2019}. 
 
Arguably, how TEs contribute to genome folding will certainly receive more attention in the future, but it is an equally fundamental question for both genome and TE biology to understand how genome folding affects TE integration preferences. For example, depending on the 3D organisation of chromosomes, a new TE copy that enters the nucleus from the cytoplasm will come across distinct parts of the genome in terms of their accessibility and organisation compared to a preexisting TE insertion that integrates into a different genomic locus. Intriguingly, retroelements (including retrotransposons and retroviruses) and DNA transposons have different replication and transposition pathways~\cite{Sultana2017}, which implies that genome architecture may have a different impact in each TE type. 

There is a clear gap in the experimental and theoretical work on understanding the impact of genome architecture on integration of TEs. Models of TE amplification dynamics have traditionally been based on population-based approaches~\cite{Charlesworth1983}, which typically set up systems of (stochastic) ordinary differential equations accounting for generic competing elements during TE expansion~\cite{Xue2016,LeRouzic2007,Roessler2018}. Few works, instead, have considered the 1D distribution of nucleosomes along the genome in order to predict preferential sites of HIV integration~\cite{Naughtin2015}. Both these classes of models necessarily neglect the multi-scale 3D organisation of the genome, i.e. from nucleosomes to TADs and from compartments to chromosome territories~\cite{Bonev2017,Cremer2001}. Because of this, they are not suited to predict the ``topography'' of TEs, i.e. the pattern of genomic sites in which TEs will preferentially integrate.

Here, we introduce and discuss a computational model based on principles of polymer physics, which aims to dissect the interplay between genome organisation and biases in TE integration. The last few years, polymer models have been proved to be very successful tools to rationalise 3D genome folding~\cite{Mirny2011,Brackley2012,Michieletto2016prx,Michieletto2017nar,Jost2014B,Bianco2018,Sanborn2015}. We first briefly review the existing framework of polymer models, then discuss a recent development of such a model to understand the physical principles of HIV integration, and finally present preliminary data obtained by extending these models to the case of TEs. We conclude by discussing potential future directions in this unexplored line of research. 

\section{Main Text}

\subsection{Biophysical Principles of Genome Folding}

While genomes are, biologically speaking, the carrier of genetic information they also are, physically speaking, long polymers. 
Polymers are well-known objects that have been studied for several decades in particular in relation to industrial applications, such as rubbers~\cite{Doi1988,Gennes1979}. Pioneers in polymer physics realised a long time ago that they obey ``universal'' laws that are independent of their chemical composition~\cite{Flory1953}. 
For instance, the way a long polystyrene molecule folds in space must be identical, statistically speaking, to that of a long DNA molecule in the same conditions. Because of this, polymer physicists typically employ ``coarse-grained'' approaches, which blur the chemical details and only retain the necessary ingredients that allow the formulation of simple and generic (universal) frameworks. Universality then implies that these coarse-grained models have predictive power for a broad range of systems with different chemistry.

\begin{figure}[b]
\centering
\includegraphics[width=0.45\textwidth]{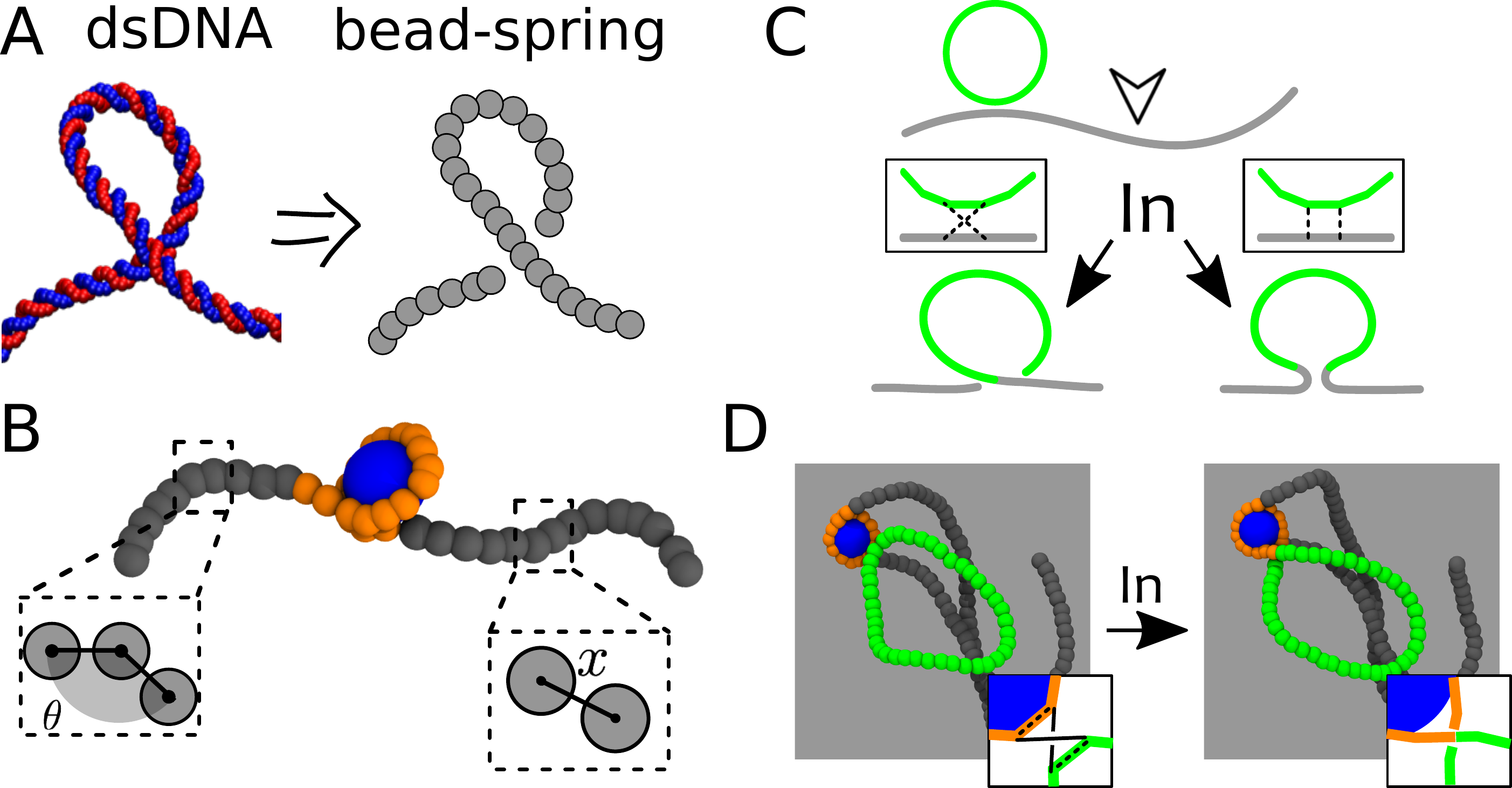}
\caption{\textbf{A} Coarse graining of microscopic details of double stranded DNA into a bead-spring polymer. \textbf{B} A polymer model for the nucleosome: highlighted are the features of DNA stiffness (set by penalising large angles $\theta$ between consecutive pair of monomers) and connectivity (set by penalising large extensions $x$ between consecutive beads). We also account for excluded volume interactions and pair-attraction represented by the wrapping of the orange segment around the histone octamer (here a blue spherical bead). \textbf{C} Schematics showing that any integration event on DNA must at least partially deform the substrate. \textbf{D} Snapshots from molecular dynamics simulations showing an integration event within a nucleosome (adapted from Ref.~\cite{Michieletto2019}). Color scheme: orange = wrapped host DNA, green= viral DNA, grey = non-wrapped host DNA.  } 
\label{fig:CG}
\end{figure}

Coarse-graining several base-pairs and groups of atoms into mesoscopic beads (see Fig.~\ref{fig:CG}A), while retaining the salient physical behaviour of DNA, allows the formulation of computational models that can reliably predict the spatial organisation of whole chromosomes from minimal input -- such as epigenetic patterns and generic binding proteins~\cite{Jost2014B,Brackley2016nar,DiPierro2016,Nuebler2018,Buckle2018,Brackley2017prl,Fudenberg2016,Sanborn2015} -- and disentangle the contribution of different classes of proteins to genome folding~\cite{Orlandini2019,Nuebler2018,Pereira2018}.

These computational models, coupled to Chromosome Conformation Capture (and its higher order variants, such as HiC) experiments~\cite{Lieberman-Aiden2009,Dekker2017}, are providing new information on the spatio-temporal organisation of the genome in different conditions, such as healthy~\cite{Brackley2016genomebiol,Buckle2018}, senescent~\cite{Chiang2018} and diseased~\cite{Bianco2018} cells, or even during cell-fate decisions~\cite{Stadhouders2019} and reprogramming~\cite{Stadhouders2018}. For instance, polymer models can rationalise features such as TADs~\cite{Benedetti2014,Fudenberg2016}, compartments~\cite{Brackley2016nar,Jost2014B} and loops~\cite{Fudenberg2016,Sanborn2015} seen in experimental HiC maps~\cite{Rao2014}. Importantly, these works are proving that traditionally physical phenomena such as liquid-liquid and polymer-polymer phase separation~\cite{Brangwynne2009,Caragine2018,Cho2018,Brangwynne2015,Michieletto2016prx,Erdel2018a}, gelation~\cite{Michieletto2019rnareview,Khanna2019}, emulsification~\cite{Hilbert2018} and viscoelasticity~\cite{Lucas2014,Shin2018} may be found in ubiquitous and key biological processes such as transcription, replication, mitosis, RNA splicing and V(D)J recombination to name a few. Polymer models are thus providing the community with a physical lens through which they may interpret complex data, and a quantitative framework to generate de novo predictions.
In light of this, we here propose that the use of polymer models may shed new light into the relationship between TE transposition and 3D organisation. 
Earlier this year Michieletto et. al. made a significant step in this direction by formalising a polymer-based model for understanding the site selection features displayed by HIV integration in the human genome~\cite{Michieletto2019}. Below, we briefly review this work, which will then be used as a stepping stone to formalise a polymer model for TE expansion.

\subsection{A Polymer Model for HIV Integration}
One of the least understood features of HIV integration is that its integration patterns display markedly non-random distributions both along the genome~\cite{Wang2007} and within the 3D nuclear environment~\cite{Marini2015}. 
HIV displays a puzzling bias for nucleosomes~\cite{Pruss1994a,Pruss1994}, gene-rich regions~\cite{Wang2007} and super-enhancer hotspots~\cite{Lucic2019} that has defied comprehension for the past three decades. Clearly, from the perspective of a retrovirus such as HIV, integrating in frequently transcribed regions is evolutionary advantageous. But how is this precise targeting achieved?

For the past decades, the working hypothesis to address this important question was that there must exist specialised factors or protein chaperones that guide HIV integration site selection. Prompted by this hypothesis, much work has been devoted to discover and identify such proteins~\cite{Kvaratskhelia2014}. Some factors, such as the lens‑epithelium‑derived growth factor~\cite{Ciuffi2005} (LEDGF/p75) have been proposed as potential candidates for this role but even knocking-down their expression could not completely remove the bias for gene-rich regions~\cite{Schrijvers2012}. Additionally, the preference of HIV to integrate in nucleosomes -- oppositely to naked DNA -- was shown in vitro using minimal reaction mixtures~\cite{Pruss1994,Benleulmi2015}. 

We recently put forward a different working hypothesis to address the bias of HIV integration site selection: could there be universal (non-system specific) physical principles that -- at least partially -- can contribute to biasing the site-selection of HIV integration? 
Prompted by this hypothesis, we decided to propose a polymer-based model in which retroviral integration occurs via a stochastic and quasi-equilibrium topological reconnection between 3D proximal polymer segments. In other words, whenever two polymer segments (one of the host and one of the invading DNA) are nearby in 3D we assign a certain probability for these two segments to reconnect, based on the difference in energy between the old and new configurations. This strategy is known as a Metropolis criterion and it satisfies detailed balance, thus ensuring that the system is sensitive to the underlying free energy~\cite{Frenkel2001}. [Note that in vitro HIV integrase works without the need of ATP~\cite{Pruss1994}, and we therefore assume that the integration process is in (or near) equilibrium.] 
Finally, we impose that the viral DNA is stuck once integrated in the selected location and cannot be excised within the simulation time. 

\begin{figure*}[t]
\centering
\includegraphics[width=0.9\textwidth]{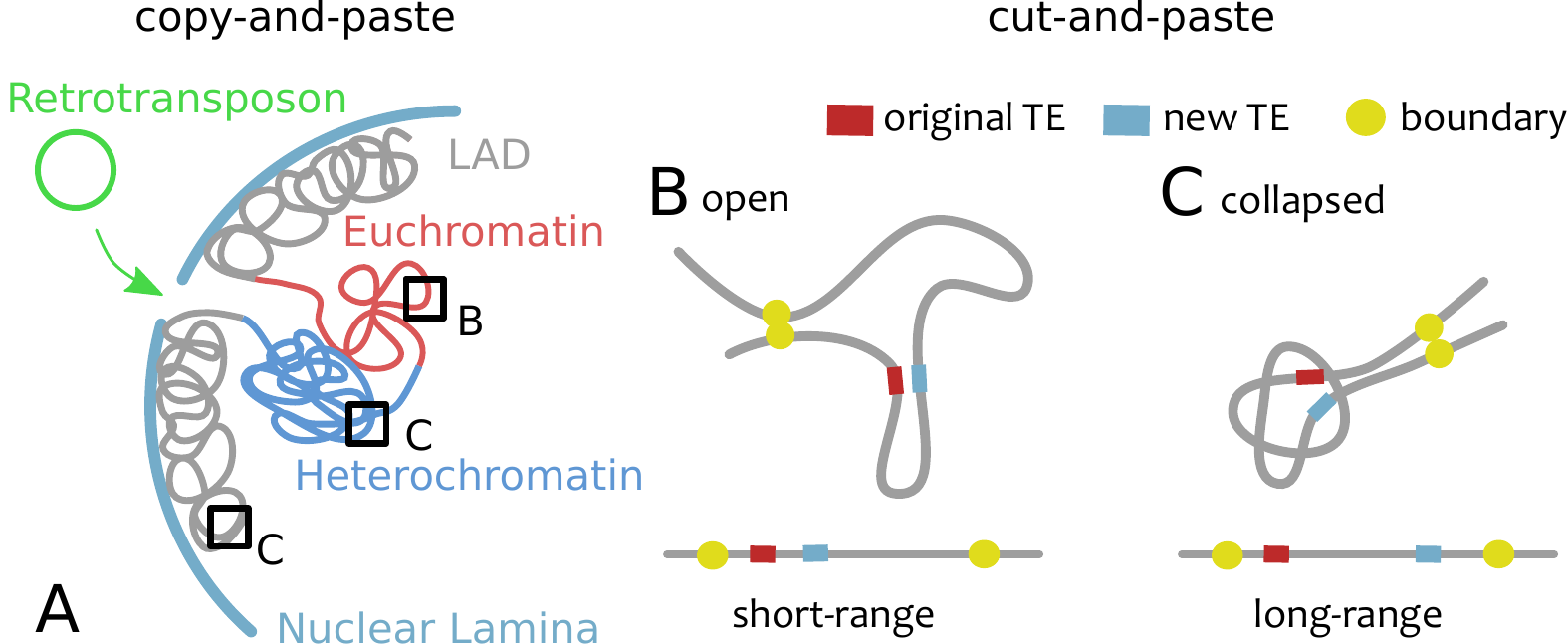}
\caption{\textbf{A} Copy-and-paste transposition explores the nuclear space by diffusing from the periphery towards the interior, i.e. outside-in. The large-scale nuclear architecture, i.e. inverted or conventional~\cite{Solovei2009}, LADs~\cite{Kind2013}, compartments~\cite{Rao2014} and enhancers hot-spots~\cite{Lucic2019,Marini2015}, are expected to play the biggest roles. \textbf{B-C} Cut-and-paste transposition explores the nuclear interior inside-out. In this case, TAD-scale ($\sim$1 Mbp) genome folding is expected to dominate and in particular open conformations will yield short range de novo re-integration whereas collapsed ones will lead to longer range re-integration.} 
\label{fig:copy_vs_cut}
\end{figure*}

Within this simple model, we discovered that geometry alone may be responsible for a bias in the integration of HIV in nucleosomes (Fig.~\ref{fig:CG}B-D). This is because the pre-bent conformation of DNA wrapped around the histone octamer lowers the energy barrier against DNA deformation required to integrate the viral DNA into the host (Fig.~\ref{fig:CG}B, see also Refs.~\cite{Serrao2014,Benleulmi2015,Pasi2016}). Further, by considering a longer region of a human chromosome folded as predicted by the above mentioned polymer models~\cite{Brackley2016nar,Michieletto2016prx,Michieletto2017nar}, we discovered that at larger scales HIV integration sites obtained from experiments~\cite{Wang2007} are predominantly determined by chromatin accessibility. Thus, by accounting for DNA elasticity and chromatin accessibility -- two universal and cell unspecific features of genome organisation -- our model could predict HIV integration patterns remarkably similar to those observed in experiments in vitro~\cite{Pruss1994,Pruss1994a} and in vivo~\cite{Wang2007}.

\subsection{Extension to DNA Transposition}
In light of the success of this polymer model, we now propose to extend it to understand the distribution of integration sites across the TE phylogeny. Importantly, different TEs have different integration strategies, which suggests that they are likely to interact differently with the 3D genome organization within the nucleus. TEs can be primarily distinguished based on the mechanism through which they proliferate, i.e. via ``copy and paste'' (class I or retrotransposons) or ``cut and paste'' (class II, DNA transposons). The former require an RNA intermediate to proliferate and thus exit the nuclear environment, whereas the latter simply relocate their DNA via endonuclease excision~\cite{Wicker2007,Sultana2017}. 

Retroviruses, like HIV, are very similar to retrotransposons with the addition that they can exit to the extracellular space and invade other cells. Thus, a new copy of a retrotransposon (class I or copy-and-paste) or a retrovirus must travel from the periphery to the nuclear interior while the genome is ``scanned'' from the outside-in for integration sites (Fig.~\ref{fig:copy_vs_cut}A). This implies that the global, nuclear-scale genome architecture is expected to be important for this re-integration process. For instance, Lamin Aassociated Domains (LAD)~\cite{Kind2013} positioning with respect to nuclear pores, inverted versus conventional organisation~\cite{Solovei2009}, compartments~\cite{Rao2014} and enhancer hot-spots~\cite{Lucic2019} will likely play major roles for retrotransposons. On the contrary, a DNA transposon (class II or cut-and-paste) probes the genome in the immediate surrounding of its excision site and will diffuse from the inside-out (Fig.~\ref{fig:copy_vs_cut}B-C). As a result of this, the mesoscale ($\sim$ 1 Mbp) organisation of the genome may have a profound effect on the 1D genomic distribution of DNA transposons. For example, heterochromatin-rich chromatin is thought to be collapsed~\cite{Boettiger2016,Michieletto2016prx} with a typical overall size that depends on the genomic length as $R \sim L^{1/3}$; on the other hand, euchromatin-rich compartments~\cite{Rao2017} are more open~\cite{Gilbert2004} and their size may be more similar to that of a random walk, i.e. scaling as $R \sim L^{1/2}$. The contact probability of two genomic loci at distance $s$ can be estimated to scale as $P_c \sim s^{- 3 \nu}$~\cite{Mirny2011} where $\nu$ is $1/3$ for collapsed polymers (such as heterochromatin), $1/2$ for ideal ones (such as euchromatin~\cite{Boettiger2016}) and $3/5$ for self-avoiding walks~\cite{Gennes1979}. Thus, a crude calculation would predict that a DNA transposon should re-integrate at distance $s$ with a probability $P_c(s) \sim s^{-3 \nu}$ that depends (through the exponent $\nu$) on the folding of the genome at these (TAD-size) length-scales. A similar effect is at play in the enhancement of long range contacts in oncogene-induced senescent cells~\cite{Chiang2019}. 

\begin{figure}
\centering
\includegraphics[width=0.47\textwidth]{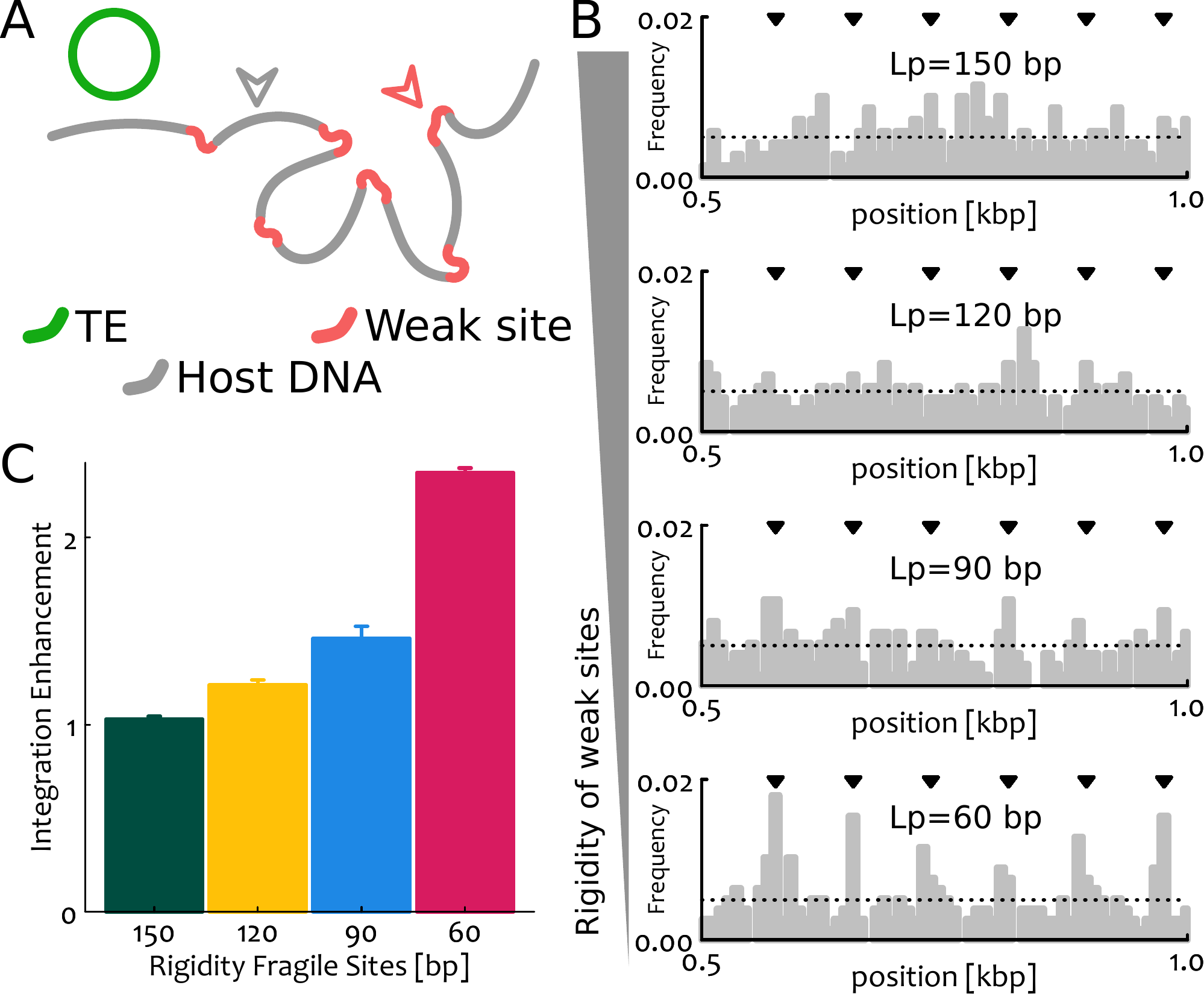}
\caption{\textbf{A} Sketch of the original simulations performed in this work where we consider a segment of DNA 1.6 kbp long (or $N=200$ beads, each bead representing 8 bp) with rigidity $l_p=150$ bp. The DNA is interspersed with ``weak'' sites which display a different rigidity $l_f$. The length of these weak sites is $8$ bp, or 1 bead.  \textbf{B} We compute the frequency of integration events per each segment of the substrate by counting the number of events occurring at a specific locus over the total integration events. We average over 1000 independent simulations. One can notice that the patterns, which are roughly uniform for $l_f=l_p$ become more and more periodic and reflecting the positions of the weak sites (denoted by the black arrows) when we reduce $l_f$. The dotted line shows the expected frequency for random events $1/N$, with $N=200$ the length of the substrate. For clarity we report only the segment 0.5-1 kbp.  \textbf{C} Integration enhancement in weak sites over the expected random frequency. Each box represents a different value of the rigidity of the weak sites $l_f$. Recall that $l_f=l_p=150$ bp reflects a uniformly stiff substrate and indeed we recover the expected value (unity) for the enhancement. 
} 
\label{fig:Integration}
\end{figure}

In addition to this contribution coming from large- and meso-scale folding, one may argue that there ought to be other complementary effects such as specific features of the integrase~\cite{Taganov2004,Serrao2014} or tethering~\cite{Gao2008} enzymes. These orthogonal elements are more local and are expected to equally affect both DNA transposons and retrotransposons. To investigate the role of local chromatin features on a generic integration event, we here performed original simulations on a heterogeneously flexible polymer that crudely mimics heterogeneous chromatin in vivo (Fig.~\ref{fig:Integration}A). Specifically, we consider a stretch of 1.6kbp long DNA with persistence length $l_p=150 \text{bp} = 50 \text{nm}$ and regularly interspersed with weak sites that display a lower bending rigidity $l_f$. This lower local DNA rigidity may be due to, for instance, to denaturation bubbles~\cite{Fosado2017}, R-loops~\cite{Santos-pereira2015} or replication stress~\cite{Chan2009}. In these conditions -- which may be reproduced \emph{in vitro} by considering DNA with a sequence of bases that modulates its local flexibility~\cite{Pruss1994a} --  we ask what is the integration pattern displayed by an invading DNA element by counting the number of integration events in each segment of the polymer over many (1000) independent simulations. We observe that, by varying the value of the rigidity parameter from $l_f=l_p=150$ bp to $l_f=60$ bp, the integration patterns become less uniform, more periodic and reflecting the distribution of weak sites (Fig.~\ref{fig:Integration}B). 

From these patterns we can compute the enhancement of integration in susceptible sites due to their different flexibility. This is simply the sum of integration frequencies in all susceptible sites divided by the one expected for a random distribution of events, i.e. $n/N$ where $n$ is the number of weak sites and $N$ the length of the polymer. This calculation is reported in Fig.~\ref{fig:Integration}C and shows that the enhancement increases with the flexibility of the susceptible sites. The output of these simulations may be readily measured in experiments in vitro on designed DNA and chromatin templates as done for HIV~\cite{Pasi2016,Benleulmi2015}, and may thus inform the mechanistic principles leading to DNA integration. 
Perhaps more importantly, however, these simulations suggest that the heterogeneity of the DNA (or chromatin) substrate in both mechanics and folding may affect TE expansion with potentially important and far-reaching consequences on the evolutionary paths and proliferative success of certain TEs in vivo. 


\section{Conclusions}
It is now becoming increasingly clear that cell function, health and fate are correlated to 3D genome folding~\cite{Bonev2017,Stadhouders2019}. TEs are intrinsically linked to 3D organisation as they are ``living elements'' within a complex multi-scale environment. In the last few years, there have been a handful of studies that started to interrogate how TEs shape genome organisation, from demarcating TAD boundaries~\cite{Zhang2019,Sun2018,Kruse2019} to harboring binding sites for architectural proteins~\cite{Choudhary2018}. It is thus now realized that TEs have profound implications in the fate and health of a cell -- not only via the traditional pathway of genomic instability and epigenetic silencing -- but also through the global regulation of genome folding. 

Now, while this crucial relationship will certainly receive more attention in the future, we argue that the other direction of the relationship, i.e. how 3D structure affects de novo TE insertions, is also of utmost importance. For example, biases in insertion patterns due to tissue-specific genome organisation in the germline (versus, for example, somatic cells) can create preferential pathways for genome evolution. We suggest that the dissection of this interplay could be done by employing a ``perturb-and-measure'' strategy of inducing TE expansion in a cell line whilst obtaining information on the 3D genome organisation and epigenetic states pre and post expansion. This approach will determine -- also through the use of polymer physics models -- which 3D and/or epigenetic features are associated with de novo TE insertions and thus detect insertion biases. Consequently, it will allow the generation of ``topographical maps'' of TE insertions in  a given tissue-specific 3D genome organisation. Ultimately, understanding the preferential insertion of TEs may lead to a better understanding of genome and TE evolution or even inform better strategies to drive genomic variations in crops. 

\section{Acknowledgements}
HWN, AB and DM acknowledge support of EPSRC and BBSRC through the Physics of Life Network under the form of Sandpit grants. HWN, DB and DM would also like to acknowledge the networking support by the ``European Topology Interdisciplinary Action'' (EUTOPIA) CA17139. HWN is supported by The Royal Society (award numbers UF160138). AB is supported by The Royal Society (award numbers UF160222 and RGF/R1/180006). 

\bibliography{library}
\end{document}